\documentclass{article}
\usepackage{amsmath,amssymb}
\usepackage{graphicx}
\usepackage[german,english]{babel}

\usepackage{oljour}
\usepackage[numbers,sort&compress,comma,square]{natbib}

\usepackage{color}

\begin{document}
\journalname{zp}

%\title[de]{Aufenthaltswahrscheinlichkeiten eines Probeteilchen unter externer  Kraft im Glas: Ergebnisse aus Modenkopplungstheorie und Simulationen aktiver   Mikrorheologie}
\title[en]{Probability densities of a forced probe particle
  in glass: results from mode coupling theory and simulations of active
  microrheology}
\begin{author}
  \anumber{1} \firstname{Ch. J.}  \surname{Harrer} \vita{}
  \institute{Fachbereich Physik, Universit\"at Konstanz} \street{} \number{}
  \zip{78457} \town{Konstanz} \country{Germany} \tel{} \fax{} \email{}
\end{author}
\begin{author}
  \anumber{2} \firstname{A. M.}  \surname{Puertas} \vita{}
  \institute{Departamento de F\'\i{}sica Aplicada, Universidad de
    Almer\'\i{}a} \street{} \number{} \zip{04.120} \town{Almer\'\i{}a}
  \country{Spain} \tel{} \fax{} \email{}
\end{author}
\begin{author}
  \anumber{3} \firstname{Th.}  \surname{Voigtmann} \vita{}
  \institute{Zukunftskolleg der Universit\"at Konstanz und Institut f\"ur
    Materialphysik im Weltraum, Deutsches Zentrum f\"ur Luft- und Raumfahrt
    (DLR)} \street{Linder H\"ohe} \number{} \zip{51170} \town{Cologne}
  \country{Germany} \tel{} \fax{} \email{}
\end{author}
\begin{author}
  \anumber{4} \firstname{M.}  \surname{Fuchs} \vita{} \institute{Fachbereich
    Physik, Universit\"at Konstanz} \street{} \number{} \zip{78457}
  \town{Konstanz} \country{Germany} \tel{} \fax{} \email{}
\end{author}
\corresponding{matthias.fuchs@uni-konstanz.de}

\def\eqref#1{(\ref{#1})}

\def\Fex{\ensuremath{F_\text{ex}}} \def\vFex{\ensuremath{\vec F_\text{ex}}}
\def\Fexc{\ensuremath{F_\text{ex}^c}} \def\kT{\ensuremath{k_\text{B}T}}
\def\mean{\ensuremath{\langle z \rangle}} \def\varz{\ensuremath{\langle \Delta
    z^2 \rangle}} \def\varx{\ensuremath{\langle \Delta x^2 \rangle}}
\def\varu{\ensuremath{\langle \delta u^2 \rangle^{\rm gl}}}
\def\skew{\ensuremath{\gamma_1}}

\abstract{ We investigate the displacements of a probe particle inside a
  glass, when a strong external force is applied to the probe (active
  nonlinear microrheology). Calculations within mode coupling theory are
  presented for glasses of hard spheres and compared to Langevin and Brownian
  dynamics simulations. Under not too strong forces where the probe remains
  trapped, the probe density distribution becomes anisotropic. It is shifted
  towards the direction of the force, develops an enhanced tail in that
  direction (signalled by a positive skewness), and exhibits different
  variances along and perpendicular to the force direction. A simple model of
  an harmonically trapped probe rationalizes the low force limit, with strong
  strain softening setting in at forces of the order of a few thermal energies
  per particle radius.  }

%\zusammenfassung{ Wir untersuchen die Dichteverteilungen eines kolloidalen Probenteilchens in einem Glas in Abh\"angigkeit von der St\"arke einer   externen Kraft, welche auf das Teilchen wirkt.  }
%\schlagwort{aktive Mikrorheologie, Glas\"ubergang, kolloidale Suspensionen}

%\dedication{To Prof.\ Matthias Ballauff on occasion of his 60th birthday}

\maketitle

\section{Introduction}

Many colloidal suspensions are composed of constituents that have typical
dimensions in the $\mu$m range. One is therefore interested in understanding
the dynamical processes occurring in these systems on mesoscopic length scales.
A recent experimental technique to obtain such information is called active
microrheology \cite{Waigh.2005,Squires.2010}: one inserts a probe particle,
similar in size to the host particles, into the system and subjects it to
controlled external driving, such as a (constant) external force. This is
conveniently realized in colloids by laser tweezers, magnetic forces, or
carefully tailored surface-chemistry reactions \cite{Erbe.2008}.  Since
scaling arguments imply that thermal fluctuations correspond to forces in the
pN range for these colloidal systems, one can easily access the strongly
nonlinear-response regime in active microrheology.  If analyzed properly, the
nonlinear response of the probe reveals a wealth of information on the
dynamics of the host fluid; in particular if the latter shows highly
collective relaxation behavior, such as at high densities close to the glass
transition \cite{Siebenbuerger.2012} or close to jamming
\cite{Chandelier.2010} .

In the fluid, a pulled probe eventually reaches a steady-state velocity,
related to the driving force by a friction coefficient $\zeta$. Linear
response demands the velocity to be proportional to the force, hence the
friction coefficient to be a constant. In active-microrheology experiments on
dense colloidal hard-sphere model suspensions \cite{Habdas.2004}, a strong
nonlinear variation of $\zeta$ with the external force $\Fex$ was found:
strong forces drastically reduce the friction experienced by the probe. The
threshold for this ``force thinning'' effect to set in is at forces much
larger than the thermal ones, i.e., $\Fex\gg\kT/a$, where $a$ is a typical
particle radius and $\kT$ is the thermal energy.  This has been related to the
cage effect that dominates the relaxation dynamics at high
density: the probe particle is trapped by transient cages of nearest
neighbors, and the force thinning threshold marks the point where
force-induced motion overrules the structural relaxation set by the decaying
cages.

The limiting case of this scenario is the response of a pulled probe inside
either a jammed state \cite{Reichhardt.2010} or an ideal glass, i.e., an
amorphous solid where the embedding particles no longer relax to thermal
equilibrium. The mode-coupling theory of the glass transition (MCT)
\cite{Goetze.2009} explains a basic mechanism behind nearest-neighbor cages
becoming permanent. The theory has recently been extended to force-driven
active microrheology in the nonlinear-response regime \cite{Gazuz.2009}. A
remarkable consequence was found: in the ideal glass, a force threshold
$\Fexc$ exists, which marks a transition, from a strictly localized probe in a
deformed amorphous solid at low forces, to a local melting leading to
delocalization and long-range motion of the probe at large forces. Experiments
and computer simulation results in the liquid have successfully been
interpreted in the theoretical framework \cite{Gazuz.2009,Gnann.2011}. In
recent work \cite{Winter.2012,Harrer.2012}, emphasis was placed on
understanding the delocalized state, in terms of friction coefficients and
fluctuations around the mean trajectory in terms of mean-squared
displacements.

In this contribution, we focus instead on the localized region, i.e., on a
probe embedded in a frozen random environment, subject to a force that is
strong enough to leave the linear response regime, but small enough to keep
the probe particle localized. This supplements earlier work studying single
particle motion in crystalline environments \cite{Ohshima.2001} to the
disordered case.  We present results of MCT describing the probability
densities of probe positions, and discuss some of their basic statistical
properties. We compare with results from molecular-dynamics and
Brownian-dynamics computer simulations.

\section{Methods}

\subsection{Theory}

Let us recall the relevant equations from the microscopic MCT for active
nonlinear microrheology \cite{Gazuz.2009}.  The theory considers the
Fourier-transformed probe-particle form factor $f^s_{\vec q}$ at wave vector
$\vec q$. It is finite for a probe particle that is trapped in the glass,
where it describes the localized probe-density distribution at (infinitely)
long times. MCT predicts a transition from a trapped to a delocalized probe
upon increasing the external force, or decreasing the particle
interactions. The latter can result from decreasing the host-fluid density,
from increasing the temperature of the system, or from making the probe
particle smaller. Schematically simplified versions of MCT have been used to
analyze such force-induced delocalization transitions in detail
\cite{Gnann.2012}, and to describe available data such as friction
coefficients, density correlation functions, or mean-squared displacements of
the probe particle \cite{Gazuz.2009,Gnann.2011,Harrer.2012}. These analyses
point out that the idealized concept of a forced but still localized probe in a
non-relaxing amorphous host solid
is very useful in understanding the qualitative nonlinearities observed for
probes pulled through dense, complex fluids.

The real-space representation of the form factor $f^s_{\vec q}$,
\begin{equation}\label{eq:fourier}
  f^s({\vec r}) = \int \frac{d{\vec q}}{(2\pi)^3}\; e^{-i{\vec q}\cdot {\vec r}}\;  f^s_{\vec q}\, ,
\end{equation}
describes the stationary probability for the center of the probe particle to
be in an infinitesimal volume element around $\vec r$, provided that it was at
the origin initially (when the constant external force $\vFex$ was switched
on). Because of particle conservation, $f^s(\vec r)$ is normalized to
unity. In a quiescent amorphous solid, spatial isotropy is assumed, which is
reflected in the fact that $f^s_{\vec q}$ depends on the wave vector only
through its modulus, $|\vec q|$.

Spatial symmetry is broken by the external force: the spherical symmetry is
reduced to a rotational one around the force axis (taken to be the $z$
axis in the following), and the distribution in
the direction of the force becomes qualitatively different. In particular, it
allows for a non-zero mean and, more generally, non-zero odd moments of the
distribution -- something forbidden by symmetry in the quiescent case.  For
later reference, recall the standard definitions of the first few moments of
$f^s(\vec r)$: the lowest nontrivial moments are the mean $\mean$ along the
force direction, the variances $\varz$ and $\varx$ in direction of and
perpendicular to the force, and the skewness of the distribution along the
force axis, $\skew$:
\begin{subequations}\label{moments}
  \begin{eqnarray}
    \mean &=& \int d{\vec r}\; z \; f^s(\vec r)\\
    \varz &=&  \int d{\vec r} \;\left( z - \mean \right)^2 \; f^s(\vec r)\\
    \varx &=&  \int d{\vec r}\; x^2 \; f^s(\vec r)\\
    \label{momentsskew}
    \skew &=& \frac{\langle \Delta z^3 \rangle}{\varz^{3/2}} =  \frac{1}{\varz^{3/2}} \; \int d{\vec r} \; \left( z - \mean \right)^3 \; f^s(\vec r)
  \end{eqnarray}
\end{subequations}

MCT decomposes the internal forces acting on the probe into density
fluctuations of the probe and its surrounding host liquid, which we take to
consist of a colloidal dispersion dissolved in a (featureless) solvent.
Because the particle interactions are translational invariant, MCT works
with plane-wave decompositions of these fluctuations, and only at the end
Fourier back-transforms to real space are taken as in Eq.~\eqref{eq:fourier}.
The form factor $f^s_{\vec q}$ appears as the long-time limit of probe-density
autocorrelation functions, that do not decay to zero for a localized particle
(corresponding to the emergence of an elastic contribution in the scattering
spectrum). Quite generically, one derives
\begin{equation}\label{eq:fq}
  f^s_{\vec q} = \frac{m^s_{\vec q}[f,f^s]}{1 + m^s_{\vec q}[f,f^s]}\,.
\end{equation}
The cage effect appears in the retarded friction and
its history-dependence. It is captured in a generalized friction kernel
$m^s_{\vec q}$ for which exact formal expressions are known that, however, can
only be evaluated with approximations. MCT approximates the host-probe force
fluctuations using effective interactions provided by the Ornstein-Zernike
direct correlation functions. As a consequence, only wavevector-dependent
density fluctuations need to be calculated within the theory. The direct
correlation functions here are assumed to be known from some separate theory,
e.g., density functional or liquid state theory \cite{Hansen}.  In the
idealized glass, the structure of the host system is then fully encapsulated
in $f_q$, the isotropic collective nonergodicity parameter (Debye-Waller
factor, in the case of a monodisperse one-component host).  The $f_q$ are
real-valued and well known from quiescent MCT calculations
\cite{Goetze.2009}.

The wavevector-dependent probe form factor $f^s_{\vec q}$, the central object
of our calculations, then is determined by the MCT closure to
Eq.~\eqref{eq:fq}, given through the mode-coupling functional
\cite{Gazuz.2009}
\begin{equation}\label{eq:mq}
  m^s_{\vec q}[f,f^s]=\int \frac{d{\vec k}}{8\pi^3}\frac{S^s(p)^2}{nS(p)}
  \frac{(\vec q\cdot\vec p)\; \omega^s_{\vec q,\vec p}}{
    (\omega^s_{\vec q,\vec q})^2}\; f_p\, f^s_{\vec k}\,. 
\end{equation}
Here, $\vec p=\vec q-\vec k$, and $p=|\vec p|$ have been introduced for
abbreviation, and the host-liquid particle density is denoted by $n$.
The term $\omega_{\vec q,\vec p}^s=[\vec q-i\vFex/\kT]\cdot\vec p$ embodies
instantaneous forces.
$S^s(p)$ and $S(p)$ are the equilibrium static structure factors given by,
respectively, the probe-host and the host-liquid interactions, and connected
to the mentioned direct correlation functions.  The latter structure factor
also is the only input into the calculation of the collective nonergodicity
parameter $f_q$ within MCT, so that all input to determine $f^s_{\vec r}$ is
now specified.

While for $\Fex=0$ the well-studied quiescent MCT follows, finite external
forces $\vFex$ affect by Newton's laws the momentum balance and thus cause a
shift in the wavevector $\vec q$ in $\omega_{\vec q,\vec p}$. Because of the
displacement of the probe from the origin, its Fourier-transformed
distribution $f^{s}_{\vec q}$ is complex. Realness of the real space
distribution, however, follows from $f^{s*}_{\vec q}=f^s_{-\vec q}$, which is
a consequence from Eqs.\eqref{eq:fq} to \eqref{eq:mq}, at least for small and
intermediate forces. There are indications from numerical solutions of the
microscopic MCT equations that at larger forces, a bifurcation
takes place to a solution that violates this requirement \cite{Harrer.2013}.
In the previous analyses involving schematic models, where no detailed spatial
information is kept, this singularity was regularized.  Here, we only consider
external forces that are smaller than the bifurcation threshold.

Choosing the interactions between particles determines the equilibrium
structure information required in Eq.~\eqref{eq:mq}, and thus completely
specifies the solutions of the MCT equations for active microrheology.  We
will discuss a system of monodisperse hard spheres with radius $a$.  The
dimensionless density is given in terms of the packing fraction,
$\varphi=\frac{4\pi}{3}na^3$, which quantifies the volume taken by the spheres
relative to the total volume. It is the only control parameter determining the
structure of the host system. For simplicity, we will also consider a probe
particle that is identical in size to the surrounding spheres. The
Percus-Yevick approximation \cite{Hansen} of the direct correlation functions
for hard spheres is used to approximate the MCT coupling coefficients: this
liquid-state theory is well established, analytically solvable, and remains
reasonable up to high densities. Within this model, MCT predicts a glass
transition to occur at a critical packing fraction $\varphi_c\approx0.516$.
We consider the window $\varphi_c\le\varphi\le0.54$ in the following.

Equations \eqref{eq:fq} and \eqref{eq:mq} are solved on a discrete lattice of
$128\times129$ points in the magnitude--angle plane, $(|\vec q|,\vartheta)$,
where $\vartheta$ is the angle formed between $\vec q$ and $\vFex$.  Angular
integration runs over the interval $[0,\pi]$ for the angle $\vartheta$.
As $f^{s}_{\vec{q}}$ is rotationally invariant around the axis of the force it
does not depend on the azimuth $\phi$. Still integration of this angle over the
interval $[0,2\pi]$ has to be performed when evaluating the mode-coupling
functional. 
Wave-vector modulus integration starts at
$2aq_\text{min}=0.001$ up to a cutoff $2aq_\text{max}=45$, $50$, $60$, or $65$
(for $\varphi=0.516$, $0.52$, $0.53$, and $0.54$); at larger $q$, both $f_q$
and $f^s_{\vec q}$ have sufficiently decayed to zero. To allow for
sufficiently smooth interpolation of $f_q$, which is pre-determined by
equations analogous to Eqs.~\eqref{eq:fq} and \eqref{eq:mq}, the latter was
calculated on a grid of $2048$ points in $|\vec q|$ (angular integration being
performed analytically in this case).  We found no qualitative dependence on
the discretization, checked by some calculations for finer and coarser
grids. However, the inverse Fourier transform required by
Eq.~\eqref{eq:fourier} is very sensitive to cutoff effects, and for the grid
used here causes tiny rapid oscillations in $f^s(\vec r)$ that cause artefacts
when calculating moments of the distribution. This is remedied by setting
$|f^s(\vec r)|$ to zero below a threshold of $2.5\,10^{-4}$.

\subsection{Simulations}

To conduct first qualitative tests of the predicted features of the density
distributions, we also performed computer simulations of active force-driven
microrheology. Two model systems were used: strongly damped molecular dynamics
for a three-dimensional system of slightly soft spheres based on a Langevin
equation, and two-dimensional Brownian dynamics simulations for hard spheres.

The Langevin dynamics simulation has been used extensively before, both in
discussing the equilibrium approach to the glass transition
\cite{Voigtmann.2004,Weysser.2010}, and for microrheology \cite{Gazuz.2009,
  Gnann.2011}; implementation details are given in these publications.  The
$3d$ system consists of $1000$ purely repulsive particles with a power-law
interaction, $V(r)\sim r^{-36}$, and interaction radii that are randomly drawn
from a flat distribution of half-width $10\%$ to avoid crystallization.
Periodic boundary conditions are used, and the box is elongated by a factor of
$8$ in the force direction to reduce finite-size effects.  The
glass-transition packing fraction of this system is known to be
$\varphi_c\approx0.595$. Here we show simulations sweeping the external force
at a fixed density of $\varphi=0.57$, accessing the (still transiently)
frozen-in component of the probe density distribution at an intermediate time
$t\approx25\sqrt{ma^2/\kT}$ ($m$ is the particle mass). For this time, the transient
self-intermediate scattering functions at $qa\approx3$ are close to their
plateau value \cite{Gazuz.2009,Gnann.2011}.
% Averaging over $300$ independent runs
Averaging over 300 simulations each consisting of about 40 independent driving
cycles was used to improve statistics.  Simulations have also been performed
at lower densities, and at $\varphi=0.62$; for the latter, the system can no
longer be equilibrated in the available computing time, but has been aged for
$t_w=25000\sqrt{ma^2/\kT}$.  Still, with this choice, considerable aging effects
cannot be ruled out for $\Fex\lesssim35\,\kT/a$, making it preferable to
obtain the density distributions from the liquid-state simulation as mentioned
above.

The Brownian dynamics simulation considers an equimolar binary mixture of hard
discs in $d=2$ dimensions, with radius ratio $1.4$. A hybrid Monte-Carlo
scheme based on event-driven dynamics \cite{Scala.2007} is used, where a
thermostat is introduced which at every integer-multiple of a 'Brownian time'
$\tau_B$ redraws all particle velocities from an identical Gaussian
distribution. This mimics Brownian motion after many $\tau_B$. Inspired by
Carpen and Brady \cite{Carpen.2005}, the external force on the probe-particle
is implemented by a shift of its Gaussian velocity distribution by the value
$\Fex/\zeta_0$ in the fixed force direction ($\zeta_0$ is the short time
friction coefficient). The glass-transition dynamics of this system has been
analyzed in detail \cite{Weysser.2011}, establishing
$\varphi_c\approx0.795$. For the reasons explained above, we chose a density
$\varphi=0.78$ and measured the probe-particle distribution function at
intermediate times. As a probe particle, a bigger disc is chosen randomly.

\section{Results}

\begin{figure}
  \includegraphics[width=.8\linewidth]{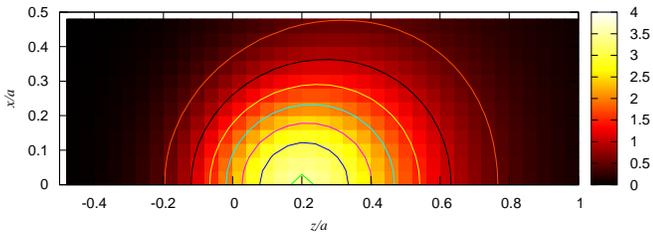}
  \caption{\label{fig:frcut} Contour plot of the probability density of a
    forced probe-sphere in a hard sphere glass at the glass transition packing
    fraction, $\varphi_c=0.516$, as calculated in MCT. The probe-particle and
    host particles are equal sized (radius $a$), and the force is $\vFex=10
    {\vec e_z} \kT/a$. A cut $f^s(x,z,y=0)$ is shown.}
\end{figure}

Figure~\ref{fig:frcut} shows a two-dimensional cut, $f^s(x,z,y=0)$, through
the probability density of the forced probe in a hard sphere glass as
calculated within MCT. The packing fraction is the critical one,
$\varphi_c=0.516$, where quiescent MCT locates the formation of an (idealized)
glass. Probe and host fluid particles are taken to be identical, which leads
to a strong coupling of the probe-particle to the collective fluid density.
For vanishing external force, the probe is localized inside nearest-neighbor
cages, and due to spatial isotropy its probability density is spherically
symmetric and centered around the origin. The width of the distribution is of
the order of a tenth of the average particle separation (that is, some
fraction of the particle diameter $2a$), as was first discussed by Lindemann
in crystals \cite{Lindemann.1910,Hansen,Goetze.2009}.

For the large force $\Fex=10\,\kT/a$ shown in Fig.~\ref{fig:frcut}, the
distribution is shifted to the wall of the cage, and a slight anisotropy
between the directions aligned with and opposite to the external force becomes
noticeable.

\begin{figure}
  \includegraphics[width=.8\linewidth]{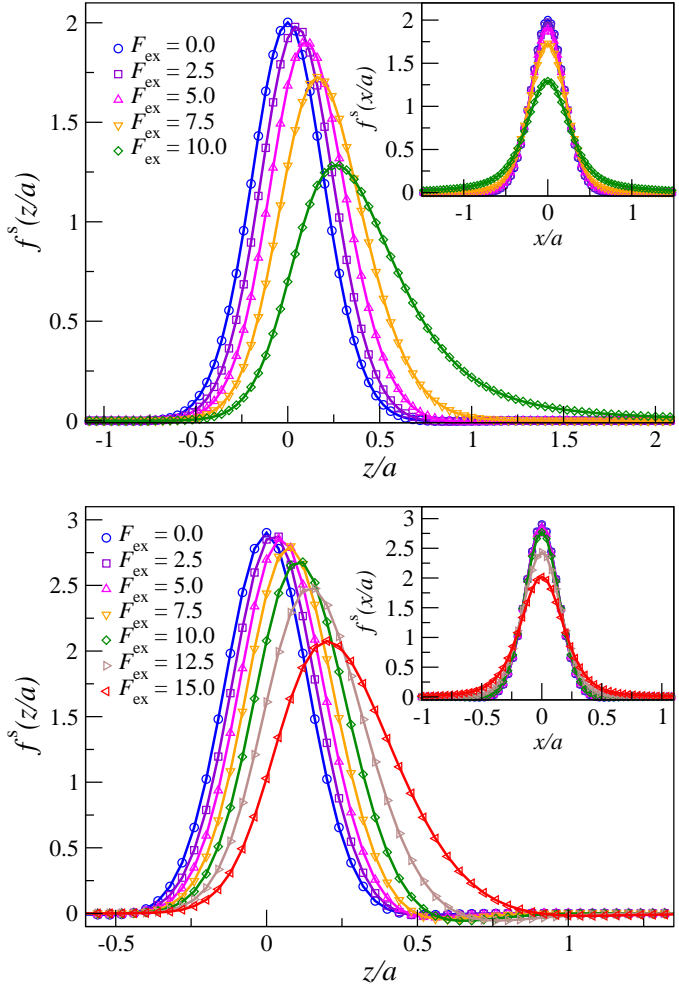}
  \caption{\label{fig:fr0516} Stationary density distributions of the
    probe-particle along the direction of the external force as calculated by
    MCT, and for forces as labeled in units of $\kT/a$. Probe and fluid
    particles have radius $a$. The upper panel is at the packing fraction of
    the MCT glass transition, $\varphi_c=0.516$, the lower one in a glass at
    $\varphi=0.53$ (note the different horizontal scales). The insets show the
    corresponding probability densities along the $x$-direction perpendicular
    to the force (with again different horizontal scales). }
\end{figure}

To identify the anisotropic effects induced by the external force more
clearly, it is convenient to discuss the one-dimensional marginal probability
densities obtained from integrating out two spatial directions.  Of particular
interest is $f^s(z)$, the probability distribution along the force axis, shown
in Fig.~\ref{fig:fr0516} for forces starting from zero up to the value
$\Fex=10\,\kT/a$ shown in Fig.~\ref{fig:frcut}, in steps of $2.5\,\kT/a$. In
the insets of Fig.~\ref{fig:fr0516}, we also show $f^s(x)\equiv f^s(y)$, the
marginal distributions along directions perpendicular to the force. We show
results for both the glass-transition density, and a state point deeper in the
glass, $\varphi=0.53$, where the probe remains localized for a larger range of
forces. The qualitative effects of the external force are the same
irrespective of density (as long as one stays inside the
glass). Quantitatively, the distributions for the higher density are narrower,
reflecting the fact that particles are more tightly localized in a more
densely packed system; note the different horizontal scales in the two panels
of Fig.~\ref{fig:fr0516}.

With increasing force, the half-width at half maximum of the distribution
increases, starting from Lindemann's value at zero force. This indicates that
cages are continuously widened by applying an increasing external force, in
all spatial directions.  The center of the distribution along the force axis,
$f^s(z)$, shifts in the direction of the force, as is intuitive. Spatial
symmetry demands that $f^s(x)$ at the same time remains centered on zero, as
verified from the figure.

In $f^s(z)$, a characteristic asymmetry develops, with an increased wing
extending in the direction of the force stronger than in the direction
opposite to it.  Interestingly, also the distributions characterizing the
probe probability densities in directions perpendicular to the force show
appreciable wings at large forces, extending to distances much larger than
Lindemann's length.

Let us remark on one other feature that can be identified for the
distributions shown in Fig.~\ref{fig:fr0516}. For some state points, slightly
negative dips can be identified in the MCT calculations. This is clearly
unphysical.  One has to keep in mind that MCT is formulated in wave-vector
space, where it is designed to guarantee a number of exact properties of
correlation functions in equilibrium \cite{Goetze.2009}. Since the theory
involves approximations, this guarantee however does not carry over to the
real-space formulation obtained from Fourier-transforming the MCT results. In
fact, small negative contributions in the predicted probability densities
(van~Hove functions) of a quiescent hard-sphere glass are known
\cite{Weysser.2010}; they are thus not a consequence of extending the theory
to the nonlinear response. Note that even the equilibrium structure
approximation we use as input for our calculations, viz.\ the Percus-Yevick
approximation for hard spheres, fails to satisfy exact positivity of the
real-space pair distribution functions (albeit at higher densities than we
consider) \cite{Hansen}.  As Fig.~\ref{fig:fr0516} shows, the unphysical
negative dips become more pronounced for higher densities, but become weaker
and finally (almost) disappear with increasing force. They remain small for
all densities $\varphi\le0.54$. In the following, we will concentrate on the
upper portions of the distribution functions where we expect MCT to make
sensible predictions.

\begin{figure}
  \includegraphics[width=.8\linewidth]{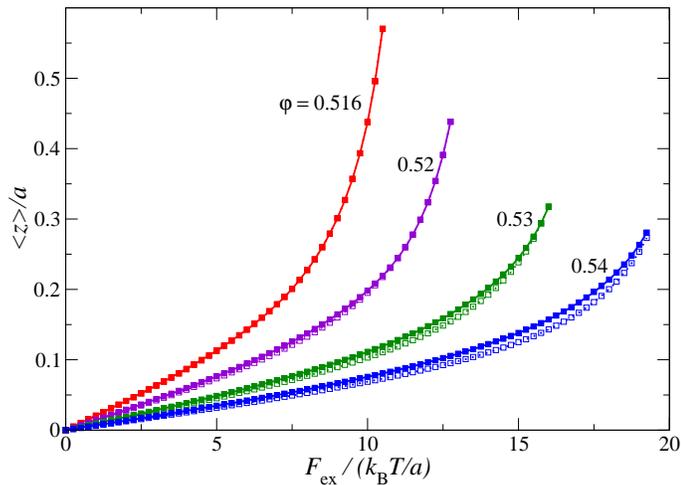}
  \caption{\label{fig:mean} Mean value $\mean$ of the long-time limit of the
    probe-particle displacement along the force direction as function of the
    external force $\Fex$. Curves for different glass packing fractions
    $\varphi$ are shown as labeled. At the higher two $\varphi$,
    corrected (filled squares) and uncorrected (open squares) $\mean$ differ
    slightly; see text for details. }
\end{figure}

From the overall shape of the distributions shown in Fig.~\ref{fig:fr0516} it
is apparent that their first few moments allow to quantify the mentioned
qualitative effects. We now turn to a discussion specifically of the mean,
variance, and skewness of $f^s(z)$ and $f^s(x)$. In calculating these
quantities, the unphysical negative portions have to be treated with care.  In
particular high-order moments tend to emphasize the small errors identified in
Fig.~\ref{fig:fr0516}. We thus introduce an ad-hoc correction by setting the
negative dips to zero when calculating moments (the distribution is
re-normalized afterwards). Values obtained from both the uncorrected and the
corrected distributions are shown in the following figures, to allow an
assessment of our ad-hoc correction.

Figure \ref{fig:mean} shows the mean values $\mean$ of the probe-displacements
in response to the external force $\vFex$ for a number of glass states. The
curves are continued up to the values of the external force where the
mentioned bifurcation in the MCT equations takes place. Beyond
this force, which depends on density, we consider the microscopic MCT results
not trustworthy. With increasing density, the shift of $\mean$ with $\Fex$
becomes smaller as the elastic strength of the glass also increases.
Initially, $\mean$ grows linearly with $\Fex$: this may be called Hookean
behavior, and is expected within the general approach of linear-response
theory. We discuss this in more detail below, in connection with
Fig.~\ref{fig:knew}. Approaching the bifurcation point, a highly nonlinear
dependence of $\mean$ on $\Fex$ is seen. Since $\mean$ grows stronger than
linearly, one may call this ``strain softening''. (Reassuringly, the effect of
the mentioned ad-hoc correction is small and only becomes visible for the
highest densities considered here.)

\begin{figure}
  \includegraphics[width=.8\linewidth]{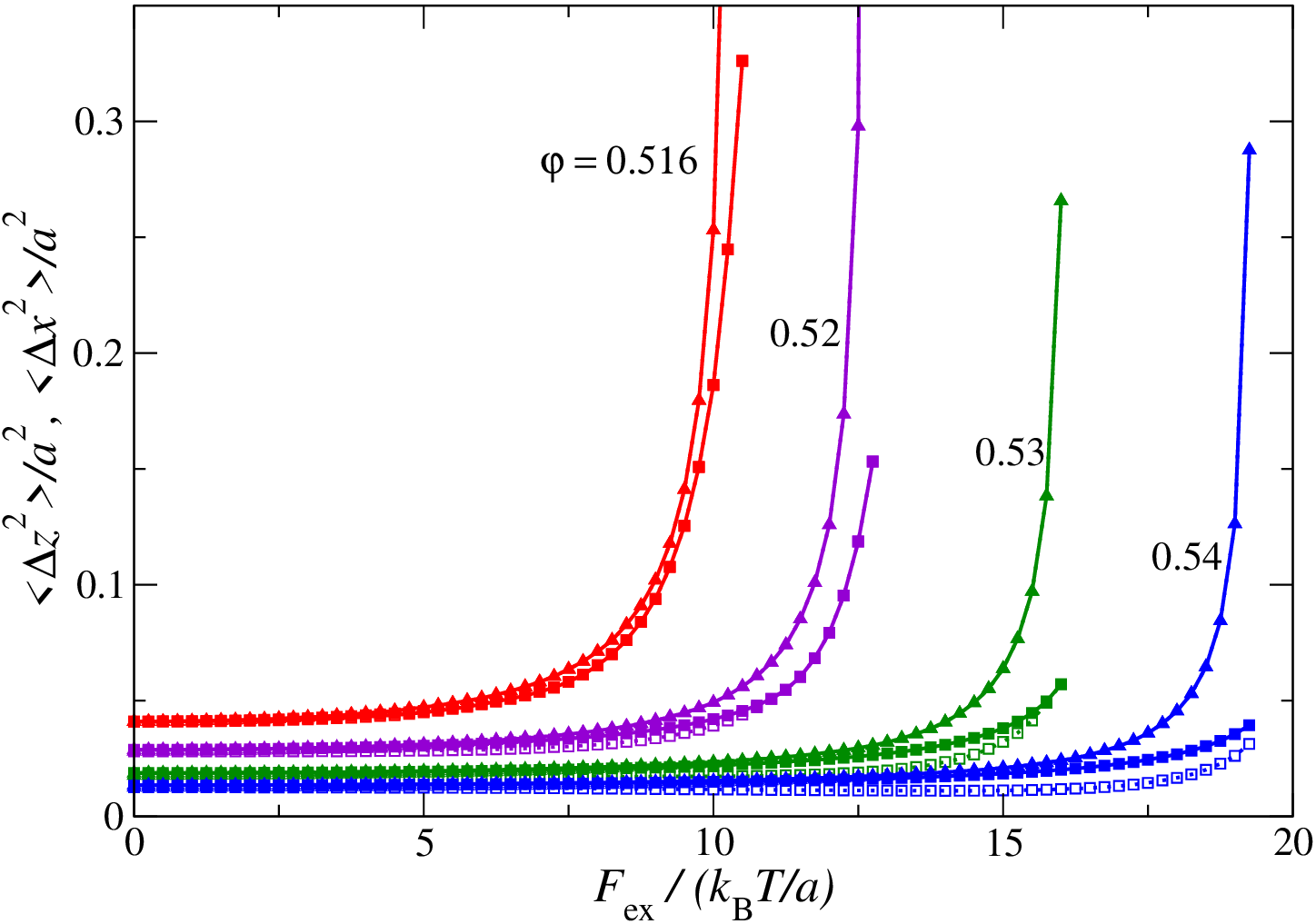}
  \caption{\label{fig:var} Variances $\varz$ (squares) and $\varx$ (triangles)
    of the stationary probe-particle density distribution along, respectively,
    perpendicular to the force direction as function of the external force
    $\Fex$. Curves for different glass packing fractions $\varphi$ are shown
    as labeled. At the higher two $\varphi$
    the corrected (filled squares) and uncorrected (open squares) $\varz$-data
    differ slightly; see text for details. }
\end{figure}

Figure \ref{fig:var} shows the variances defined in Eq.~\eqref{moments} as
functions of the strength of the external force for the four glass densities
used above.
Again, corrected and uncorrected values agree quite well except for small
differences in $\varz$ at the highest densities. At small $\Fex$, the
probe-particle response remains nearly isotropic, as required by linear
response. Both $\varz$ and $\varx$ approach for $\Fex\to0$
values given by the localization
length of the quiescent glass, $r_s$. This quantity has been evaluated before
\cite{Goetze.2009} and has been connected to the Lindemann criterion
already mentioned: the localization length $r_s$ of single particles inside a
crystal cannot exceed some fraction of the lattice constant, as long as the
crystal is supposed to be stable against thermal fluctuations
\cite{Lindemann.1910}. MCT transfers this concept to a disordered solid, where
the average particle separation plays the role of the relevant length
scale. The value $\varz=2r_s^2\approx0.04a^2$ at the transition density
$\varphi_c$ has already been verified by dynamic light scattering experiments
on hard-sphere-model colloidal supsensions \cite{Megen.1998,Sperl.2005} and by
simulations \cite{Voigtmann.2004}.  In these cases, the time-dependent
mean-squared displacement $\langle\delta r^2(t)\rangle$ was studied, whose
long-time limit in the glass is $\langle\delta r^2(t)\rangle=3\varz$ assuming
spatial isotropy.

For increasing force $\Fex$, Fig.~\ref{fig:var} confirms the qualitative
judgement of Fig.~\ref{fig:fr0516}: the probe-particle distributions become
wider in all directions, revealing a rather steep increase as the bifurcation
force is approached from below. Outside the linear-response regime, the
response also becomes anisotropic. The distribution perpendicular to the force
direction becomes wider more quickly than along the force. We attribute this
effect to the wide tails seen in the insets of Fig.~\ref{fig:fr0516}.

\begin{figure}
  \includegraphics[width=.8\linewidth]{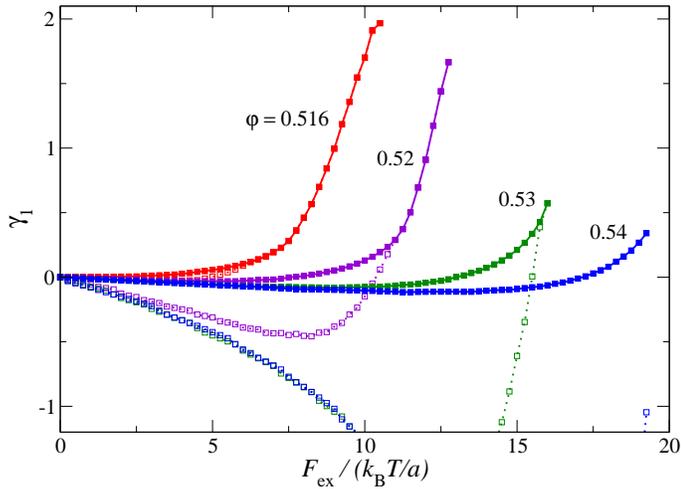}
  \caption{\label{fig:skewness} Skewness $\skew$ of the probe-particle density
    distribution along the force direction as function of the external force
    $\Fex$. Curves for different glass packing fractions
    $\varphi$ are shown as labeled. At all $\varphi$ except for
    the lowest one, corrected (filled squares) and uncorrected (open
    squares) $\skew$ differ, because the latter are affected by the (small)
    negative tails in $f^s(z)$ predicted by MCT. Thus, we judge the corrected
    ones more reliable; see text for details. }
\end{figure}

We next discuss the skewness of the distributions, Fig.~\ref{fig:skewness}.
Here, the small negative dips in the MCT approximation to the real-space
distribution functions are amplified so that the uncorrected values of
$\gamma_1$ become strongly negative for large densities and strong forces. The
ad-hoc corrected values remain small at small forces and then are positive and
increase appreciably for larger $\Fex$; since we expect MCT to work best for
the central features of the distribution we consider these latter values to be
more accurate. Generically, the marginal distribution $f^s(z)$ is hence
right-skewed, i.e., its mass is concentrated to values lower than the mean,
indicating that a few realizations correspond to large excursions in the
direction of the force.  This is most prominent close to the bifurcation
point, where again the most steep increase is seen.

\section{Discussion}

\subsection{Preliminary comparison with simulations}

In order to compare with computer simulation, we face the problem that
the idealized glass state considered by MCT cannot be obtained in the
simulation: at high densities, the system can no longer be equilibrated
in the available computer time, and aging effects prevail that are outside
the scope of our discussion. We hence choose a density close to the MCT
glass transition, where the random host structure still relaxes in the
simulation time window. Before this final relaxation sets in, the intermediate
nearly-frozen structure can be probed.

\begin{figure}
  \includegraphics[width=.9\linewidth]{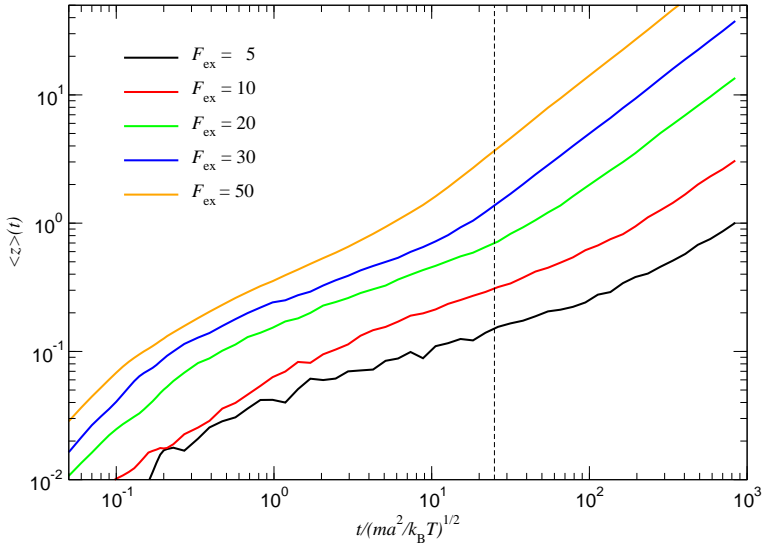}
  \caption{\label{fig:zsim} Linear displacement of the forced tracer particle in
    direction of the force (given in $\kT/a$) for the density
    $\varphi=0.57$ from Langevin dynamics simulations. The intermediate time
    $t=25\sqrt{ma^{2}/\kT}$ at which probe-particle distributions were sampled
    is indicated  by the dashed vertical line. }
\end{figure}

To demonstrate this, Fig.~\ref{fig:zsim} shows the linear displacement
$\mean(t)$ of the probe particle as a function of time since switching on
the external force, obtained from the Langevin-dynamics simulation.
One recognizes that these functions grow linearly with time at large $t$,
indicating fluid-like steady-state motion corresponding to a finite friction
coefficient $\zeta$. This steady state is reached on a time scale governed
by the host-fluid relaxation at small forces. Before that, $\mean(t)$
exhibits a shoulder as a function of $\log t$, indicative of transient
localization. We pick an intermediate time $t=25\sqrt{ma^2/\kT}$ (indicated
by a dashed line in Fig.~\ref{fig:zsim}) where probe-particle distributions
are sampled. At this time, the probe particle has fully explored its cage,
but is still sufficiently close to its transient localized state for a range of low
forces, roughly $\Fex\le 20 \kT/a$. For the higher forces, the time when the probe is pulled out-off the cage moves to shorter values, and the probability distribution sampled at $t=25\sqrt{ma^2/\kT}$ becomes characteristic for the delocalized state.  For $\Fex > 20 \kT/a$ at the given packing fraction ($\varphi=0.57$) the simulation results for $f^s$ thus will contain different phenomena than discussed from the theory.

\begin{figure}
  \centerline{\includegraphics[width=1\linewidth]{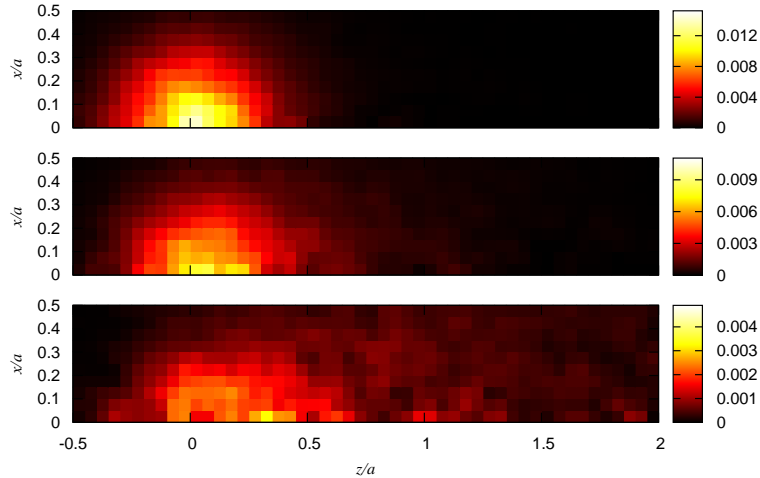}}
  \caption{\label{fig:cage} Probability density of a forced probe in a
    polydisperse hard sphere fluid at $\varphi=0.57$ from Langevin dynamics
    simulations. From top to bottom: $\Fex/(\kT/a)=$ $10$, $20$, and $30$; here
    the force is along the $x$-axis. The colour code for the unnormalized
    distribution is given in the upper panel. The time since switching-on the
    force is chosen such that the localized density distribution is close to
    its frozen-in form at the glass transition.}
\end{figure}

Figure~\ref{fig:cage} shows the probe-particle distribution $f^s(\vec r)$
sampled at that time, for various external forces.
As in Fig.~\ref{fig:fr0516}, a cut in a plane
containing $\Fex$ has been chosen, and rotational symmetry around that
axis has been used to improve statistics. The panels correspond to increasing external force. One notices the
nearly intact spherical symmetry for the lowest force, $\Fex=10\,\kT/a$, which
is broken for the larger forces.  Qualitatively, one recognizes the features
discussed above from the MCT predictions: an increasing shift of the center of
the distribution in the direction of the applied force, and an increase in the
variance in all spatial directions.

\begin{figure}
\centerline{  \includegraphics[width=.8\linewidth]{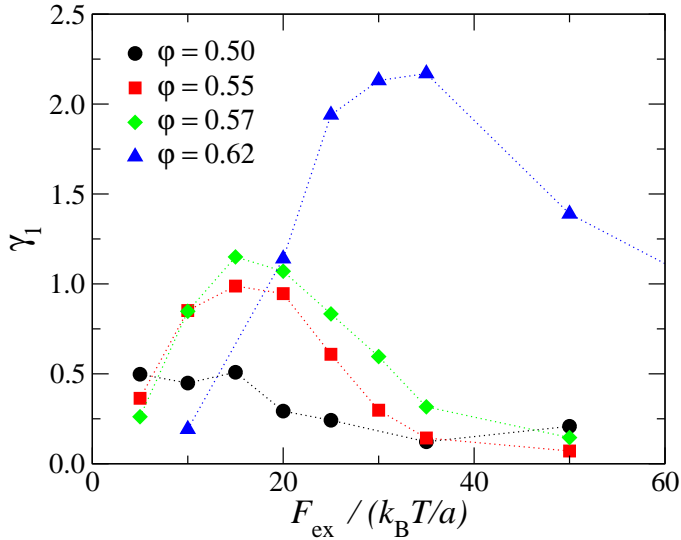}}
  \caption{\label{fig:skew} Langevin simulation results for the skewness
    $\skew$ of the probability distribution along the force-direction as a
    function of the force, for different densities as labeled. Forces are
    given in units of $\kT/a$.}
\end{figure}

The heterogeneity of the probe dynamics is parametrized by the skewness of
these distributions in Fig.~\ref{fig:skew}, as defined in
Eq.~\eqref{momentsskew}. Increasing the external force from zero, the skewness
increases, before it reaches a maximum at intermediate forces and finally
decays to zero again. From the above discussion of Fig.~\ref{fig:zsim} and a comparison with the mean probe velocity
\cite{Gazuz.2009}, we associate this decay with the eventual delocalization of
the probe; this effect is not contained for the small forces discussed within
MCT above. There is qualitative agreement between theory and simulation for
the initial increase: The skewness is positive, indicating that the probe
undergoes rare excursions far in the direction of the force, i.e., the density
distribution develops a tail there. While in the fluid, the maximum of $\skew$
lies at the same force, in the glass the increase shifts to larger forces.
This is compatible with the MCT scenario of a critical bifurcation force
whose value in the fluid close to the glass transition is given by the
one at the transition.
Values of $\skew\approx1$ at forces around $\Fex\approx10\,\kT/a$ at the glass
transition density are found in MCT; this is in semi-quantitative agreement
with the simulations.

\begin{figure}
  \includegraphics[width=.9\linewidth]{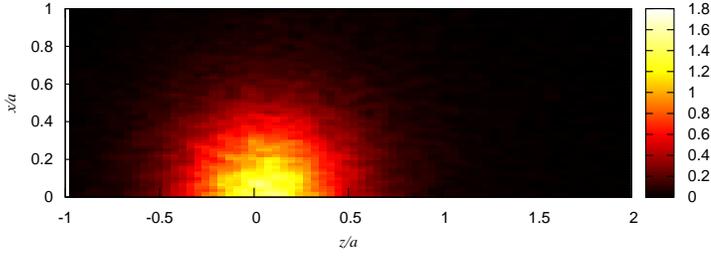}
  \caption{\label{fig:frsim} Probability density of a forced probe in a binary
    hard disk mixture (radius ratio $a_b/a_s=1.4$) in $d=2$-dimensions close
    to the glass transition density from Brownian dynamics simulations. The
    probe has been chosen among the bigger disks,  the applied force is
    $\Fex= 42\kT/a$ in $z$-direction and $a=0.6$ denotes the average particle
    radius. The colour code gives the
    probability. The time since switching-on the force is chosen such, that
    the localized density distribution is close to its frozen-in form at the
    glass transition.}
\end{figure}

Also the two-dimensional Brownian dynamics simulations of active
microrheology, broadly give results in agreement with the theoretical
predictions. Figure \ref{fig:frsim} shows a contour plot of the density
distribution of a probe (one of the larger species) trapped in an equimolar
mixture of disks with radii-ratio $a_b/a_s=1.4$. The density is close to the
glass transition one, and a large force $\Fex=42\kT/a$ is applied. Here $a$
denotes the average particle radius $a=(a_b+a_s)/2$ .
Estimates of the low-order moments, $\mean\approx0.14a$, 
$\varz\approx0.32a^2$, $\varx\approx0.24a^2$, and $\skew\approx2.51$ lead
to values which can be taken as extrapolations of the theoretical trends to
such strong forces; the small average displacement is noteworthy though, and
presumably due to the lower spatial dimension. The asymmetry of the density
distribution along the force direction is apparent already from the contour
plot.

\subsection{Small force model of an harmonically trapped probe}

\begin{figure}
  \includegraphics[width=.99\linewidth]{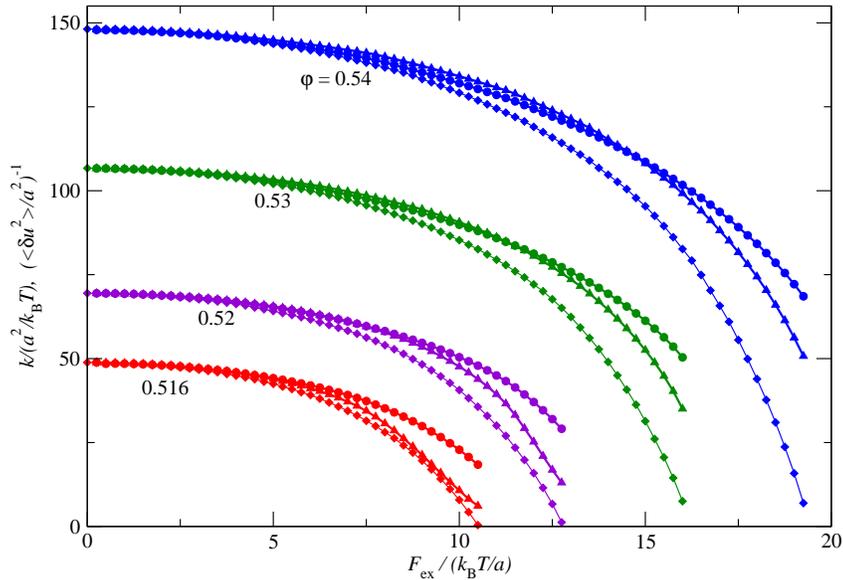}
  \caption{\label{fig:knew} Effective force constant $k$ (circles) determined
    from the probe-particle displacement under force for different glass
    packing fractions $\varphi$ as listed in the legend. Testing the linear
    response relation Eq.~\eqref{linear}, the displacement variances along
    $\kT/\langle \delta u_z^2\rangle^{\rm gl}=2\kT/\varz$ (triangles) and
    $\kT/\langle \delta u_x^2\rangle^{\rm gl}=2\kT/\varx$ (squares)
    perpendicular to the force are compared to $k=\Fex/\langle \delta u
    \rangle=\Fex/\mean$. For small forces, where $ \varu\to r_s^2$
    isotropically, linear response and the equipartition theorem hold.  Shown
    are only the corrected values not affected by the (small) negative tails
    in $f^s(z)$ predicted by MCT; see text for details.}
\end{figure}

To place the findings reported above in theoretical context, we briefly
discuss the linear-response theory of a forced particle in a glass.  A
prototypical model system is that of a Brownian particle in a harmonic
trap. Considering for simplicity only the force direction, there holds the
Langevin equation
\begin{equation}\label{hosci1}
  \zeta \dot{u}(t) = - k \;(u(t) - u_0) + \Fex + f(t)\;.
\end{equation}
Here, $u(t)$ is the displacement variable of the particle, and $k$ the spring
constant of the harmonic potential. Random forces $f$ are assumed to be
Gaussian white noise obeying the fluctuation-dissipation theorem connected to
the friction coefficient $\zeta$. The minimum position of the potential is
given by $u_0$. The distance from this minimum, $\delta u=u(t)-u_0$
immediately gives the linear shift and the equal-time variance of the
displacement at infinite times,
\begin{equation}\label{linear}
  \langle \delta u \rangle^{\rm gl} = \frac{\Fex}{k} \; ,\qquad 
  \left.\varu\right|_{\Fex=0} = \frac{\kT}{k}\; .
\end{equation}
Yet, the glass state is not an equilibrium state, but a metastable state,
where ergodicity has been broken. The nonergodicity parameters $f_q$ of MCT
quantify this, and a finite localization length $r_s=\sqrt{\langle \delta
  z^2(t\to\infty)\rangle/2}$ is the simplest example of a nonergodic quantity
in the probe dynamics. Importantly, the localization length is measured in the
limit of infinite times, where the maximal excursion $\delta z(t) = z(t) -
z(0)$ of the tracer can be obtained.
%Glass is nonergodic, as the mean squared
%displacement does not increase diffusively at long times, which holds for the
%ergodic fluid state.
The question arises how to define ergodic, close
to (metastable) equilibrium fluctuation functions in the glass state, which
are assumed in the simple harmonic oscillator model.  Recently, in the context
of measuring strain fluctuations in colloidal glass by microscopy, this
problem has been solved \cite{Klix.2012}.  The stationary (quasi-equilibrium)
displacement auto-correlation function in the glass state is defined from the
mean squared displacement defined in the fluid state by
\begin{equation}\label{glass1}
  \langle \delta u(t) \delta u(0) \rangle^{\rm gl} = \frac 12 \left( \varz - \langle \left( z(t) - z(0) \right)^2 \rangle \right)\;.
\end{equation}
Here averages on the right are performed using the Gibbs-Boltzmann
distribution of the fluid state, while the average on the left hand side is
within the compartment in phase space corresponding to the nonergodic glass
state. The function $\langle \delta u(t) \delta u(0) \rangle^{\rm gl}$ defined
on the left hand side obeys all the properties required of an auto-correlation
function, at least if the approximate MCT equations of motion are taken as
basis \cite{Klix.2012}. Clearly and as expected, Eq.~\eqref{glass1} gives a
linear relation between the displacement $u(t)$ of the probe in the glass and
the position $z(t)$, whose long time distribution functions were the topic of
our MCT calculations: $u(t) = z(t)$. Somewhat more surprisingly, the second
moment of the displacement taken at equal time follows to be half as big as
the variance of the density distribution remaining at long times:
\begin{equation}\label{glass2}
  \varu=\langle \left(\delta u(t=0)\right)^2 \rangle^{\rm gl} = \frac 12 \; \varz = \frac 12 \; \langle \left( z(t\to\infty) - z(0) \right)^2 \rangle \;.
\end{equation}

Figure \ref{fig:knew} tests the thus obtained simple model of the motion of
the trapped probe in glass. The values of the effective spring constant,
$k=\Fex/\mean$, are compared to the appropriately inverted mean squared
displacements $\varu$, namely $\kT/\varu$, and quantitative agreement is found.
This supports the above definition of a displacement fluctuation function in
the nonergodic glass state. Its behavior at small forces reduces to the simple
model of an equilibrium harmonic oscillator. At larger external forces, force
thinning sets in, and the first and second displacement moments behave
qualitatively similar, yet quantitatively differ.  (In order to eliminate
spurious non-monotonic behavior arising from the negative probabilities MCT would predict,
only the corrected moments, where the negative probabilities were set to zero,
are shown in Fig.~\ref{fig:knew}.)

The linear-response spring constant evaluated by MCT close to the glass
transition is $k\approx50\,\kT/a^2$, which agrees quite well with a recent
estimate $k\approx80\,\kT/a^2$ obtained employing a parabolic trap in the
Langevin simulations \cite{Puertas.2010}. Following basic elasticity theory,
it can be related to a (local) Young's modulus $E$ of the glass. Another
relevant length enters this conversion (the cross-section over which the force
is applied divided by the original length of the strained object). Assuming
this to be roughly $a$ again, we get $E_\text{micro}\approx50\,\kT/a^3$. In
the linear macroscopic rheology of glasses, MCT predicts both the longitudinal
and the shear modulus, $M$ and $G$; for a hard-sphere system at the glass
transition, $G\approx20\,\kT/(2a)^3$ and $M=M_0+\delta
M\approx130\,\kT/(2a)^3$, where $\delta M\approx55\,\kT/(2a)^3$ is the
nonergodic contribution \cite{Goetze.2003}. From the standard relation between
the elastic moduli, $E=(G/M)(1-(4/3)(G/M))/(1-(G/M))$, one gets
$E\approx56\,\kT/a^3$, which compares reasonably well with the microscopic
spring constant.

\section{Conclusions}

We investigated the nonlinear response of a single particle subject to a
strong external force in an amorphous solid. Calculations based on the
mode-coupling theory for active nonlinear microrheology allow to access the
density distribution functions describing the average probe position in the
nonequilibrium system. These distribution functions become asymmetric and
loose the spherical symmetry present in the spatially homogeneous equilibrium
state: quite intuitively, the center of the distribution is shifted in the
direction of the applied force. In addition, two nontrivial effects are
predicted: the distribution is widened in all spatial directions, i.e., its
localization also perpendicular to the applied force becomes less strong when
the force is increased. Further, one identifies strong tails in the
distribution function, and in particular a pronounced tail extending along the
force from the initial positions, as manifested in a positive skewness
parameter. This tail becomes stronger as the force approaches a critical
bifurcation threshold. It quantifies dynamical heterogeneities that arise because
occasionally particle trajectories follow realizations with large excursions.
In fact, it is seen in the simulation that the probe-particle dynamics close
to delocalization becomes intermittent. The connection of the strong tails
and the bifurcation force found in the MCT equations is currently under
investigation.

For small forces, linear response theory allows to understand the results if
one takes into account the broken ergodicity of the glass. This proper
definition of quasi-equilibrium displacement correlation functions in glass
accounts for an otherwise puzzling factor $2$ connecting the variance of the
probe-distribution functions to the harmonic spring constant characterizing
the local rigidity of the nearest-neighbor cage.  Semi-quantitatively, the
local measure of rigidity can be connected to the macroscopic elastic
constants of the glass.

{ We thank for funding by the Deutsche Forschungsgemeinschaft throuh SFB
  Transregio TR6 and Research Unit FOR 1394. Th.~V.\ thanks for funding
  through the Helmholtz-Forschungsgemeinschaft (Impuls- und Vernetzungsfonds,
  VH-NG-406), and through the Zukunftskolleg der Universit\"at Konstanz. 
  A.M.P. acknowledges financial support from Junta de Andalucia and FEDER,
  and Ministerio de Ciencia e Innovaci\'on under projects P09- FQM-4938 and
  MAT2011-28385, respectively.
 }

\bibliography{lit} \bibliographystyle{iopart-num}

\end{document}